# Avoided Crossing Patterns and Spectral Gaps of Surface Plasmon Modes in Gold Nano-Structures

**Alexandre Kolomenskii\*, Siying Peng, Jeshurun Hembd, Andrei Kolomenski,**

**John Noel[1], Winfried Teizer and Hans Schuessler**

*Department of Physics, Texas A&M University, College Station, TX 77843-4242*

[1]*Currently with the Institute for Environmental Medicine, University of Pennsylvania, Philadelphia, PA 19104-6068*

\*Corresponding author: <a-kolomenski@physics.tamu.edu>

**Abstract**

The transmission of ultrashort (7 fs) broadband laser pulses through periodic gold nano-structures is studied. The distribution of the transmitted light intensity over wavelength and angle shows an efficient coupling of the incident p-polarized light to two counter-propagating surface plasmon (SP) modes. As a result of the mode interaction, the avoided crossing patterns exhibit energy and momentum gaps, which depend on the configuration of the nano-structure and the wavelength. Variations of the widths of the SP resonances and an abrupt change of the mode interaction in the vicinity of the avoided crossing region are observed. These features are explained by the model of two coupled modes and a coupling change due to switching from the high frequency dark mode to the low frequency bright mode for increasing wavelength of the excitation light.

PACS numbers: 73.20.Mf, 42.70.Qs, 42.25.-p,

The interaction of light with periodic metal structures, allowing for efficient coupling of light to surface plasmons (SPs) with their remarkable properties continues attracting considerable interest. In particular, sharp resonances [1-4], localization and enhancement of the electromagnetic field [5,6], and wave guiding along metal surfaces [7] were demonstrated. Corrugated metal surfaces and arrays of nanowires find applications in sensing [8] and enhancing light absorption [9,10]. Recently, after the discovery of the extraordinary light transmission through arrays of small holes [11] the investigation of the role of SPs in this phenomenon demonstrated that the transmission minima correspond to the excitation of the SP modes [12].

The interaction of light with periodic structures is also of interest, because excited SPs exhibit gaps in the energy spectrum ($\omega$-gaps) [3,5,7,13], rendering these structures as simple plasmonic crystals. The gaps appear due to mode interaction near the crossing of their unperturbed dispersion curves and can be employed for wave guiding [14], developing sensors [15] and photonic notch filters [16]. The existence of momentum gaps (or $k$-gaps) was also proposed and confirmed in experimental [13] and theoretical works [17-19]. However, it has also been suggested that a $k$-gap is an artifact, resulting from the mode over-damping [20].

In this Letter we investigate the interaction of SP modes observed with the transmission far field spectroscopy [21,22] of broadband laser pulses (transform limited duration 7 fs) through metal nano-structures and use a rather general approach of the model of coupled modes for interpretation. The incident light efficiently couples to two counter-propagating surface plasmon modes, exhibiting avoided crossing with features that were not reported previously. The experimentally measured minima in the transmitted light form patterns, indicating the existence of $\omega$- and $k$-gaps, confirmed also by calculations with the coupled mode model [23,24].

We studied two types of metallic structures: a periodic array of gold nanowires on a glass substrate and a similar array with an additional gold sub-layer. The periods of the arrays were chosen to provide SP resonances at normal incidence close to the middle of the spectral range (from 650 to 850 nm) of the laser pulses. The configurations of the two samples produced by thermal vapor deposition and electron beam lithography are shown in Fig. 1(a). Both samples had a gold grating structure with a rectangular profile. For the first sample the grating was thinner, and it was deposited onto an underlying gold film. The thickness of the films during the deposition was monitored with a quartz crystal monitor. The samples were also characterized with an atomic force microscope (AFM) and by scanning electron microscopy (SEM). A SEM image of the surface of the second sample is shown in Fig. 1(b).

SPs can be efficiently excited when the incident light interacting with the grating produces a polarization wave propagating along the metal surface with the phase velocity of the SP wave. This condition, following also from the conservation of energy and the momentum component along the surface, can be realized by adjusting the wavelength, the incidence angle of the light wave, the period of the grating [25,26] or the refractive index of the dielectric medium adjacent to the metal [27]. Thus, the resonance coupling takes place, when the wave vector of light after interaction with the grating has a component in the plane of the grating equal to the wave vector of the SP,

$$k \sin\theta + nk_{gr} = k_{sp} , \qquad (1)$$

where $k = 2\pi/\lambda$ is the wave number of light, $\theta$ is the incidence angle (see Fig.1(a)), integer number $n$ is the order of the diffraction interaction with the grating, $k_{gr} = 2\pi/d$ and $k_{sp}$ is the SP wave number.

Measurements of the transmitted light for a set of incidence angles were performed. A portion of the beam from a broadband pulsed laser (Rainbow, Femtolasers) was weakly focused and front-illuminated the sample. The sample was mounted on a rotation stage allowing for variation of the incidence angle from -6° to +6° with small steps and accuracy better than 0.05°. The light that interacted with the grating on the sample was selected by a small aperture and directed into an Ocean Optics spectrometer, registering the distribution of the spectral intensity. Normalized transmission spectra for the two samples are shown as density plots in Fig. 2(a,c). The intensity distribution over incidence angle and wavelength displays two diagonal valleys corresponding to $n=\pm 1$ orders of the SP resonance excitation, producing a decrease in the transmitted intensity. The higher frequency branch demonstrates a narrowing for sample 1 and a slight broadening for sample 2, when approaching the crossing region. This can be seen by comparing the shorter-wavelength dips of the spectral intensity profiles shown for a set of angles in Fig. 2(b,d), which exhibit asymmetric Fano-type resonances [28]. Such changes indicate damping variations of the SP modes. In particular, radiative damping can change due to a shift of the mode intensity distribution relative to the grooves of the grating, as the wavelength changes [29], and also the effect of the mode interaction due to Bragg scattering increases closer to the normal incidence, corresponding in $k_{sp}$-space to the boundary of the Brillouin zone [21].

We note that the coupling of the incident light to SPs at a grating surface is highly dependent on the polarization. The maximum coupling is achieved, when the projection of the electric field vector on the sample surface is perpendicular to the grooves. We measured SP coupling with light at normal incidence for varied polarizations, and the normalized data (not presented) suggest a $\cos^2(\varphi)$ dependence on the azimuthal angle $\varphi$, i.e. only the component with the polarization perpendicular to the grooves couples to SPs.

As was shown previously [25,26], in the optical spectral region the coupling reaches maximum values for the grating thickness $h$ from 30 to 50 nm, close to the values used for our samples. The relative strength of coupling for the lower and higher frequency branches depends also on the relative phase of the fundamental and second harmonics of the grating spatial profile [20]. The coupling was inferred from the normalized intensity distributions and characterized by the ratio of the depth of the intensity minimum to the height of the nearby maximum for observed Fano-type resonances as a function of the incidence angle (see Fig. 3). The coupling for the lower frequency branch ("bright" mode) is substantially stronger than coupling for the higher frequency branch ("dark" mode). Note that the coupling of the former increases, reaching 80%, and the coupling of the latter abruptly decreases, as the incidence angle approaches zero. For sample 2 the coupling dependences were qualitatively similar, except that the drop of the coupling for the higher frequency branch near the normal incidence was not observed.

The plots of the minima of the experimentally measured transmission show qualitatively different patterns of the dispersion relations for samples 1,2, depicted in Fig. 4(a,d). For sample 1 an $\omega$-gap of about 30 meV is observed (Fig. 4(a)), which appears as a result of the interaction of $n = \pm 1$ modes and is mainly due to the presence of the second harmonic in the spatial profile of the nano-structure [20,26]. The lower and higher frequency branches can be related to the symmetric and antisymmetric modes [20,29,30]. The edges of the gap at $\theta = 0^o$ have wavelength values $\lambda_{h.f.} = 723$ nm and $\lambda_{l.f.} = 736$ nm and two different SP phase velocities $v_{sp,h.f.}$ and $v_{sp,l.f.}$ of the higher and lower frequency SP branches, which have the following ratios to the speed of light $c$ in air, $v_{sp,h.f.}/c = d/\lambda_{h.f.} = 0.968$ and $v_{sp,l.f.}/c = d/\lambda_{l.f.} = 0.951$, inferred from Eq. (1) for normal incidence.

For sample 2, the mode pattern in the band-gap region is qualitatively different for higher and lower frequency branches. For the higher frequency branch, the dispersion curves with $n = \pm 1$ do not tend to merge when they approach the avoided crossing region, but rather form a gap in $\theta$ values ($k$-gap, $\Delta k = 1.91 \times 10^3$ cm$^{-1}$). The origin of such a gap cannot be attributed to a partial overlap of the approaching modes in the avoided crossing region, as was suggested [20], since only *one*, namely the lower frequency branch passes the center of this region.

Figures 4(b,e) show the evolution of the transmission angular dependence for the two samples in the avoided crossing regions, indicated by dashed boxes in Fig. 4(a,d), as the wavelength is changed in small steps. With increasing wavelength the two side minima in the angular intensity distribution (Fig. 4(b), sample 1), which are seen clearly for $\lambda = 716$ nm, initially merge into a valley and then re-appear again. In Fig. 4(e) (sample 2) with increasing wavelength the two side minima stay separated until they disappear (four lower curves), and then two shallow minima are formed.

To describe the observed behavior we used a model of two coupled SP modes [23,24] with n=+1 and n=-1, counter-propagating along the z-axis and coupled with coefficients $K_{1,2}$ and $K_{2,1}$. The main mechanism of the interaction of these modes is their Bragg scattering with $2k_{gr}$ [5,20,21]. Following general approach [23,24], we find wave numbers as functions of $\omega$ taking into account the coupling of the modes:

$$k'_{sp;1,2} = \pm(k_{sp;1} + k_{sp;2} + 2k_{gr})/2 \mp q \text{ for } \omega < \omega_c \text{ and}$$

$$k'_{sp;1,2} = \pm(k_{sp;1} + k_{sp;2} + 2k_{gr})/2 \pm q \text{ for } \omega > \omega_c, \qquad (2)$$

where $k_{sp;1} = k_{sp;1}(\omega)$ and $k_{sp;2} = k_{sp;2}(\omega) = -k_{sp;1}$ are wave numbers of the two SP modes without interaction, corresponding to the incident light frequency $\omega = 2\pi c/\lambda$, and $\omega_c$ is the crossing frequency, namely the solution of the equation $k_{sp;1}(\omega_c) - k_{sp;2}(\omega_c) - 2k_{gr} = 0$;

$q = \sqrt{\Delta^2 + G}$, $\Delta = (k_{sp;1} - k_{sp;2} - 2k_{gr})/2$ and $G = K_{1,2}K_{2,1}$. These equations show that for real values of $k_{sp;1}, k_{sp;2}$ and $K_{1,2} = -K_{2,1}$ (typical for small losses and contra-directional coupling) the condition $G < -\Delta^2$ is fulfilled for an interval of $\omega$ values ($\omega$-gap), where $q$ becomes imaginary, giving rise to the avoided crossing and a strong suppression (attenuation) of the SP modes within the gap. Thus, for $G < 0$ the two SP modes form an $\omega$-gap (conservative coupling), while for $G > 0$ the gap appears in the $k'_{sp}$ values (dissipative coupling) [24]. When attenuation of the SP modes takes place or the coupling constants are complex, the values of $k'_{sp;1,2}$ also become complex, so that the real parts $\text{Re}[k'_{sp;1,2}]$ determine the phase constants of the modes, and the imaginary parts $\text{Im}[k'_{sp;1,2}]$ determine their attenuation or amplification, depending on the sign.

The results of the dispersion relation calculations with the model of two coupled SP modes (Eq. 2) are shown in Figs. 4(c,f) as $\lambda(\theta)$ dependences to enable direct comparison with the experimental data. The unperturbed dispersion relations $k_{sp;1}(\omega)$ and $k_{sp;2}(\omega)$ were calculated using the three layer model [31], taking into account the grating as an additional gold layer with effective thickness $h^{(1)}_{eff,1} = 15$ nm and $h^{(2)}_{eff,1} = 35$ nm for samples 1 and 2, respectively. For sample 1, agreement with the experiment is obtained when the value of $G$ has a negative real part and a relatively smaller imaginary part, namely $G = (-310 - 137i)\,(\text{meV})^2$, so that in the assumption $|K_{1,2}| = |K_{2,1}| = K$ we obtain for the magnitude of the coupling constant $K = 18.4$ meV. A significant negative real part $\text{Re}[G]$ is required for a large $\omega$-gap with a strong attenuation of SP modes within the respective frequency interval. When Re[$G$] is positive, a $k$-gap (or a gap in $\theta$ values) and two almost vertical portions of the dispersion curves appear. Note that the presence of the imaginary part Im[$G$]$\neq 0$ can also lead to the formation of a $k$-gap. This is the reason for the appearance of a relatively narrow $k$-gap in calculated dispersion curves of Fig. 4(c). When this $k$-gap is small and damping is present, as in our case, the vertical portions of the dispersion curves merge, forming a vertical line of minima in the transmission, as observed in the experiment (Fig. 4(a), the data points shown by crosses). Thus, for Re[$G$]$<0$ and Im[$G$]$\neq 0$ features of both $\omega$- and $k$-gaps can be present as in Fig. 4(c), where strong attenuation of the SP modes takes place for the vertical portions of the dispersion curves.

For sample 2, the lower frequency branch looks similar to that of sample 1. However, the higher frequency branch shows the divergence of the dispersion curves for $n=+1$ and $n=-1$ in the vicinity of the avoided crossing region ($\lambda < \lambda_c = 2\pi c/\omega_c = 728$ nm), while in the region $\lambda > \lambda_c$ the portions of the dispersion curve tend to merge as the wavelength is approaching the center of the gap $\lambda_c$. The different behavior in these two wavelength regions provides evidence that the sign of the real part of the product $G$ changes, as the wavelength changes from $\lambda < \lambda_c$ to $\lambda > \lambda_c$. The solid lines in Fig. 4(f) are calculated with $G = 289$ (meV)$^2$ for $\lambda < \lambda_c$ and with $G = -576$ (meV)$^2$ for $\lambda > \lambda_c$ and give good agreement with experimental points.

The change of the coupling constant can be related to a spatial shift of the intensity distribution relative to the grating slits with increasing wavelength, as the excitation switches from the higher frequency dark mode to the lower frequency bright mode, which was recently observed in a grating structure [29]. The coupled mode model *directly calculates dispersion* *dependences of the modes*, and thus, plots of Fig. 4(c,f) clearly demonstrate the existence of $k$-gaps. The experiment shows the presence of the transmission minima also in the gap region (in Fig. 4(d) these points are shown by crosses), which are not reproduced by the presented theory. These minima gradually disappear farther away from the respective dispersion branch, which indicates their transitional nature.

In conclusion, we experimentally investigated the interaction of light with periodic gold nano-structures on glass substrates and observed qualitatively different behavior near the avoided crossing region, depending on the configuration of the sample. The sample with a grating deposited on top of a gold film exhibited an energy (or $\omega$-) gap between the two branches of the dispersion relation. The valley in the intensity distribution of the transmitted light in the vicinity of normal incidence and the calculation with the coupled mode model indicate possible presence of a small momentum (or $k$-) gap, as well. Additionally, the higher frequency branch experienced narrowing of the width of the transmission minimum and a strong reduction of the coupling to the incident light, which indicate a decrease of damping in the avoided crossing region. For the sample with only a gold grating deposited on a glass substrate the two branches of the dispersion relation were qualitatively different: while the lower frequency branch exhibited behavior typical for an $\omega$-gap, the higher frequency branch was split into two portions showing a significant $k$-gap. The coupled mode model shows that $\omega$-gaps are formed, when the coupling constants are real, and $k$-gaps appear for predominantly imaginary coupling constants. Thus, when the latter has both real and imaginary parts, features characteristic for gaps of both types are possible.

We thank Wonmuk Hwang for providing the AFM for sample profile measurements. This work was partially supported by the Robert A. Welch Foundation (grants Nos. A1546 and A1585), the National Science Foundation (grants Nos. 0722800 and 0555568) and the Air Force Office of Scientific Research (grant FA9550-07-1-0069).

**Figure captions**

FIG. 1 (color online). Measured samples: (a). Schematic of the sample profile and the geometry of the laser beam incidence (red arrows). The two samples had the following dimensions. Sample 1: $d = 705$ nm, $d_1 = 390$ nm, $h = 27$ nm, $h_1 = 35$ nm. Sample 2: $d = 700$ nm, $d_1 = 365$ nm, $h = 50$ nm, $h_1 = 0$. (b). SEM image of sample 2.

FIG. 2 (color online). Transmitted light intensity for different incidence angles and wavelengths: (a,b) sample 1, (c,d) sample 2. Figures (a, c) show false color density plots with the dark (blue) color corresponding to the reduction of the light transmission due to the SP excitation. Figures (b,d) depict the wavelength dependences of the intensity for a set of angles, showing that while in (b) the minimum at shorter wavelength becomes narrower for smaller angles, in (d) the similar minimum becomes slightly broader; in (b,d) the curves for different wavelengths are equidistantly shifted vertically for better viewing.

FIG. 3 (color online). Angular dependences of the coupling for the lower and higher frequency SP branches for sample 1.

FIG. 4. (color online). Observed and calculated SP modes in the avoided crossing region: (a, d) minima of the transmission from the experiment, (b,e) transmitted intensity vs. angle for a set of wavelengths and (c,f) calculation of the SP modes of $n = \pm 1$ orders taking into account their interaction (Eq. (2), details in the text). Crosses in (a,d) show the extension of the observed minima into the gap region. Dotted lines in (c,f) show the dispersion dependences of the SP modes without interaction. Transmission dependences in the band-gap regions, indicated by dashed boxes in (a,d) are plotted for different wavelengths with steps of 4 nm in (b) and with steps of 2 nm in (e) for samples 1 and 2, respectively.

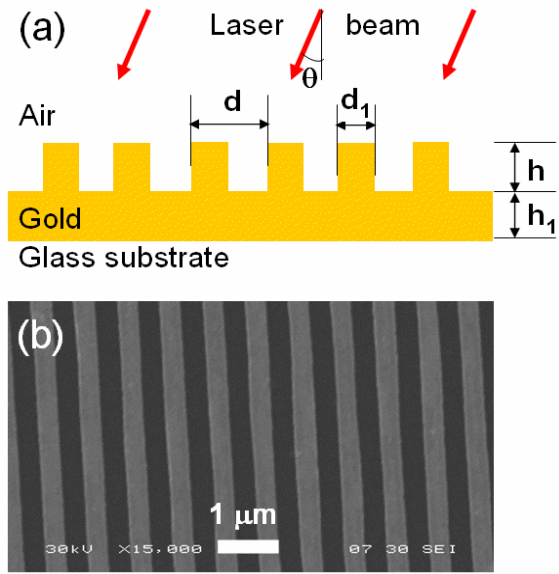

FIG. 1 (color online). Measured samples: (a). Schematic of the sample profile and the geometry of the laser beam incidence (red arrows). The two samples had the following dimensions. Sample 1: $d = 705$ nm, $d_1 = 390$ nm, $h = 27$ nm, $h_1 = 35$ nm. Sample 2: $d = 700$ nm, $d_1 = 365$ nm, $h = 50$ nm, $h_1 = 0$. (b). SEM image of sample 2.

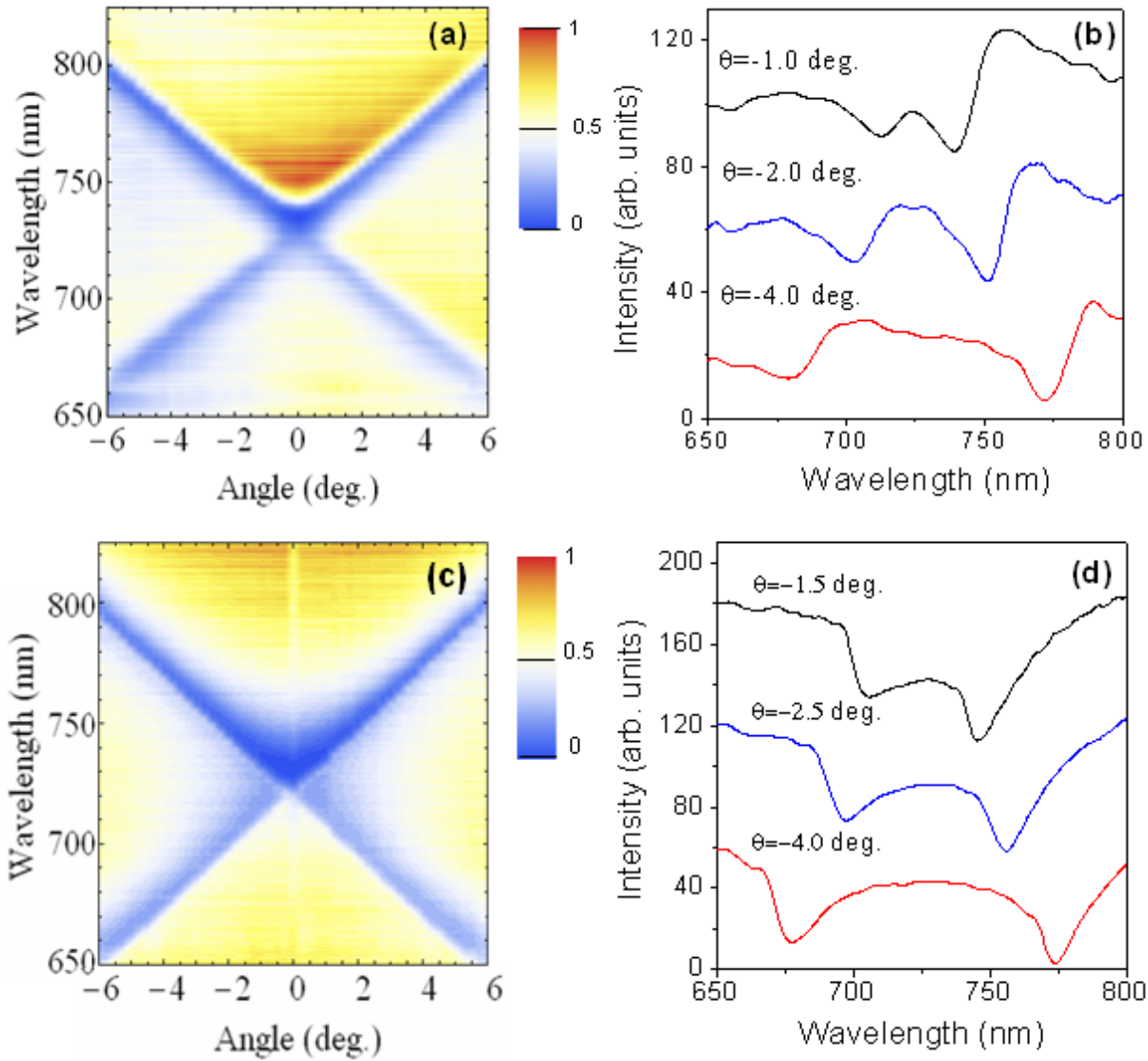

FIG. 2 (color online). Transmitted light intensity for different incidence angles and wavelengths: (a,b) sample 1, (c,d) sample 2. Figures (a, c) show false color density plots with the dark (blue) color corresponding to the reduction of the light transmission due to the SP excitation. Figures (b,d) depict the wavelength dependences of the intensity for a set of angles, showing that while in (b) the minimum at shorter wavelength becomes narrower for smaller angles, in (d) the similar minimum becomes slightly broader; in (b,d) the curves for different wavelengths are equidistantly shifted vertically for better viewing.

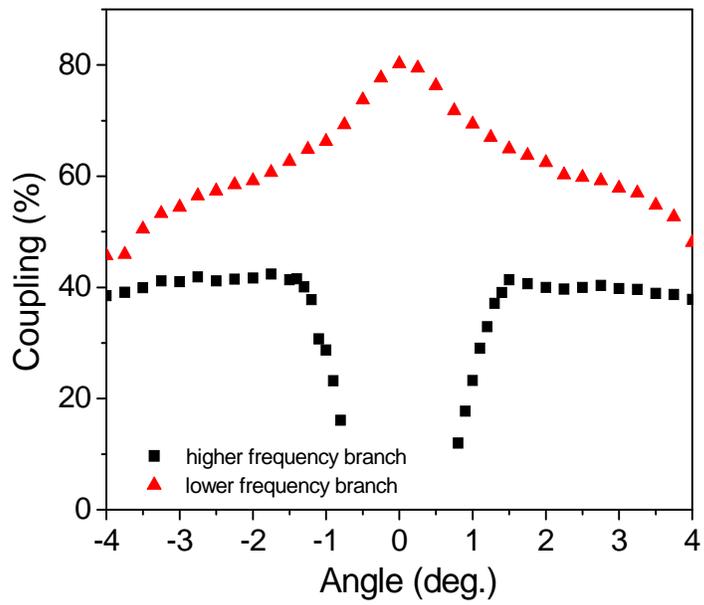

FIG. 3 (color online). Angular dependences of the coupling for the lower and higher frequency SP branches for sample 1.

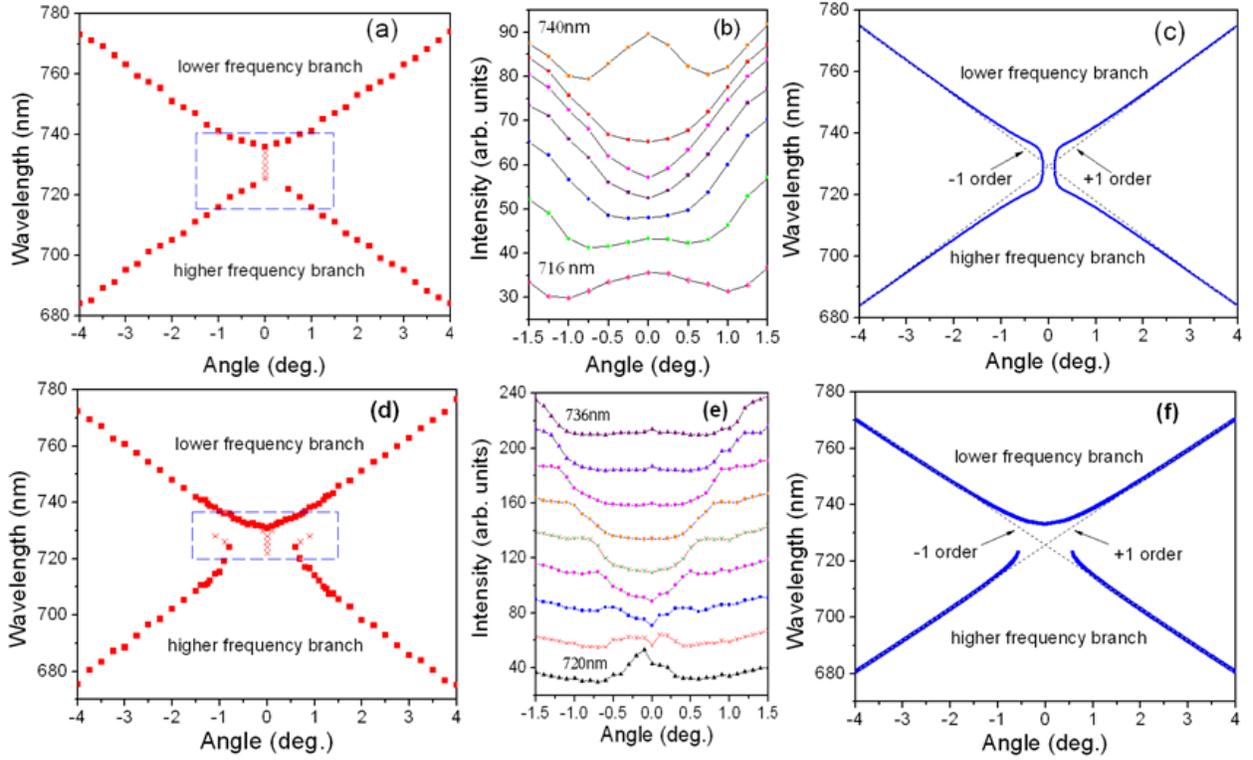

FIG. 4 (color online). Observed and calculated SP modes in the avoided crossing region: (a, d) minima of the transmission from the experiment, (b,e) transmitted intensity vs. angle for a set of wavelengths and (c,f) calculation of the SP modes of $n=\pm 1$ orders taking into account their interaction (Eq. (2), details in the text). Crosses in (a,d) show the extension of the observed minima into the gap region. Dotted lines in (c,f) show the dispersion dependences of the SP modes without interaction. Transmission dependences in the band-gap regions, indicated by dashed boxes in (a,d) are plotted for different wavelengths with steps of 4 nm in (b) and with steps of 2 nm in (e) for samples 1 and 2, respectively.